\documentclass[11pt]{article}
\usepackage[a4paper,left=0.5in,right=0.5in,top=0.5in,bottom=0.5in]{geometry}
\usepackage{graphicx}
\usepackage{ulem}
\usepackage{amssymb}
\usepackage{amsmath}
\usepackage{multirow}
\usepackage{color}
\usepackage{ulem}
\usepackage{xfrac}
\usepackage{hyperref}
\usepackage[colorinlistoftodos]{todonotes}
\usepackage{dsfont}
\usepackage{empheq,etoolbox}
\usepackage{subcaption}
\usepackage{accents}
\usepackage{xcolor}
\usepackage[shortlabels]{enumitem}
\usepackage{url}
 \usepackage{wasysym}
\newcommand{\g}{\mathbf{g}}
\newcommand{\dd}{\mathrm{d}}

\newcommand{\ee}{{\rm e}}
\newcommand{\ii}{{\rm i}}
\newcommand{\HH}{\mathcal{H}}

\author{Manolis Perrot\footnote{Univ Lyon, Ens de Lyon, Univ Claude Bernard, CNRS, Laboratoire de Physique, F-69342 Lyon, France, contacts: pierre.delplace@ens-lyon.fr, antoine.venaille@ens-lyon.fr}, Pierre Delplace$^*$, Antoine Venaille$^*$}

\title{Topological transition in stratified \textcolor{black}{fluids}}

\begin{document}

\maketitle

\textbf{Lamb waves are trapped acoustic-gravity waves that propagate energy over great distances along a solid boundary in density stratified, compressible fluids \cite{lamb1911atmospheric,vallis2017atmospheric}. They constitute useful indicators of explosions in planetary atmospheres \cite{bretherton1969lamb,lindzen1972lamb}. When the density stratification exceeds a threshold, or when the impermeability condition at the boundary is relaxed, atmospheric Lamb waves suddenly disappear \cite{iga2001transition}. 
Here we use topological arguments to predict the possible existence of new trapped Lamb-like waves in the absence of a solid boundary, depending on the stratification profile. The topological origin of the Lamb-like waves is emphasized by relating their existence to two-band crossing points carrying opposite Chern numbers. The existence of these band crossings coincides with a restoration of the vertical mirror symmetry that is in general broken by gravity. From this perspective, Lamb-like waves also bear strong similarities with boundary modes encountered in quantum valley Hall effect \cite{xiao2007Valley,li2011topological,zhang2013valley} %\cite{xiao2007Valley,li2011topological,zhang2013valley}
and its classical analogues \cite{pal2017edge,noh2018observation,qian2018theory}.
Our study shows that the presence of Lamb-like waves encode essential information on the underlying stratification profile in astrophysical and geophysical flows, which is often poorly constrained by observations.}

\bigskip

The simplest flow model supporting acoustic and internal gravity waves involves a compressible fluid in the presence of gravity $-g \mathbf{e}_z$ in a vertical half plane $(x,z)$. Owing to compressibility, such fluids support propagation of acoustic waves with sound speed $c_s$. Gravity breaks the flow isotropy, adds an intrinsic frequency $g/c_s$ and allows for the propagation of internal gravity waves when the fluid is stratified, with density profile $\rho_0(z)$. Due to stratification, the system admits another intrinsic frequency $N=\sqrt{-g\partial_z \rho_0 /\rho_0 -g^2/c_s^2}$. This is the natural buoyancy frequency of fluid particles oscillating in the vertical direction, commonly called Brunt-V\"ais\"al\"a frequency, and known to rule a variety of phenomena in atmospheres and oceans \cite{vallis2017atmospheric}. 

\textcolor{black}{The atmospheric Lamb wave is an additional mode that is known to occur in nearly isothermal atmospheres, in the presence of a  solid horizontal boundary. This mode is peculiar, as it is vertically trapped above the ground, hydrostatically balanced,  and propagates horizontally as a non-stratified acoustic-wave. Most remarkably, the Lamb waves transit from the internal gravity wave band to the acoustic wave band when the horizontal wavenumber is varied.  The existence of an edge state that transits between two wave bands describing bulk eigenmodes (not localized on the boundary) is usually reminiscent of topological waves that are currently being discovered in virtually all fields of physics. The topological nature of such boundary modes is revealed through their robustness against various kinds of perturbations, and in particular to any modifications of the boundary. From this standpoint, the robustness of the Lamb wave is unsound as it was shown by Iga that this mode can disappear when changing the boundary condition \cite{iga2001transition}. 
 We show in the following that topological Lamb-like waves exist in the canonical model for compressible stratified fluids in the absence of boundary, depending on the stratification profile.}

% The dynamics of stratified compressible flow models involves four fields: 
The dynamics of stratified compressible flow models involves the horizontal and vertical velocity components $(u,w)$, the potential temperature $\theta$ and the pressure $p$. The temporal evolution of those fields is given by the conservation of momentum in both directions, mass and entropy. 
 After proper rescaling of the different fields (see supplementary material 1), the linear dynamics around a state of rest with stable stratification (i.e. $N^2>0$) is conveniently expressed as
\begin{align}\label{eq:transf}
\partial_t\left(\begin{array}{c}
{u}\\ {w} \\ {\theta} \\ {p}
\end{array}\right)=
\left(\begin{array}{cccc}
0 & 0 &0&-c_s\partial_x\\
0&0&-N&S-c_s\partial_z\\
0&N&0&0\\
-c_s\partial_x&-S-c_s\partial_z&0&0
\end{array}\right)\left(\begin{array}{c}
{u}\\{w} \\ {\theta} \\ {p}
\end{array}\right)\ ,
\end{align}
where the \textit{stratification parameter} 
\begin{equation}
S=\frac{1}{2}\left(\frac{N^2c_s}{g}-\frac{g}{c_s}\right) 
\end{equation}
plays a central role, as it breaks mirror symmetry in the $z$-direction $(z,w,\theta) \rightarrow -(z,w,\theta)$.

Textbooks usually focus on the particular case of an ideal gas with an isothermal background stratification. In that case thermodynamical constraints impose $S<0$ (see supplementary material). Here we will relax this simplifying assumption by considering the possibility for $S \geqslant 0$. This can for instance be achieved by considering non-isothermal atmospheres with sufficiently large buoyancy frequencies $N$.  In the remaining of this paper, we adimensionalize the equations using $c_s^2/g$ and $c_s/g$ as a length and time unit, respectively, so that  $S=\frac{1}{2}(N^2-1)$.

\begin{figure}[h!]
\centering
\includegraphics[width=18cm]{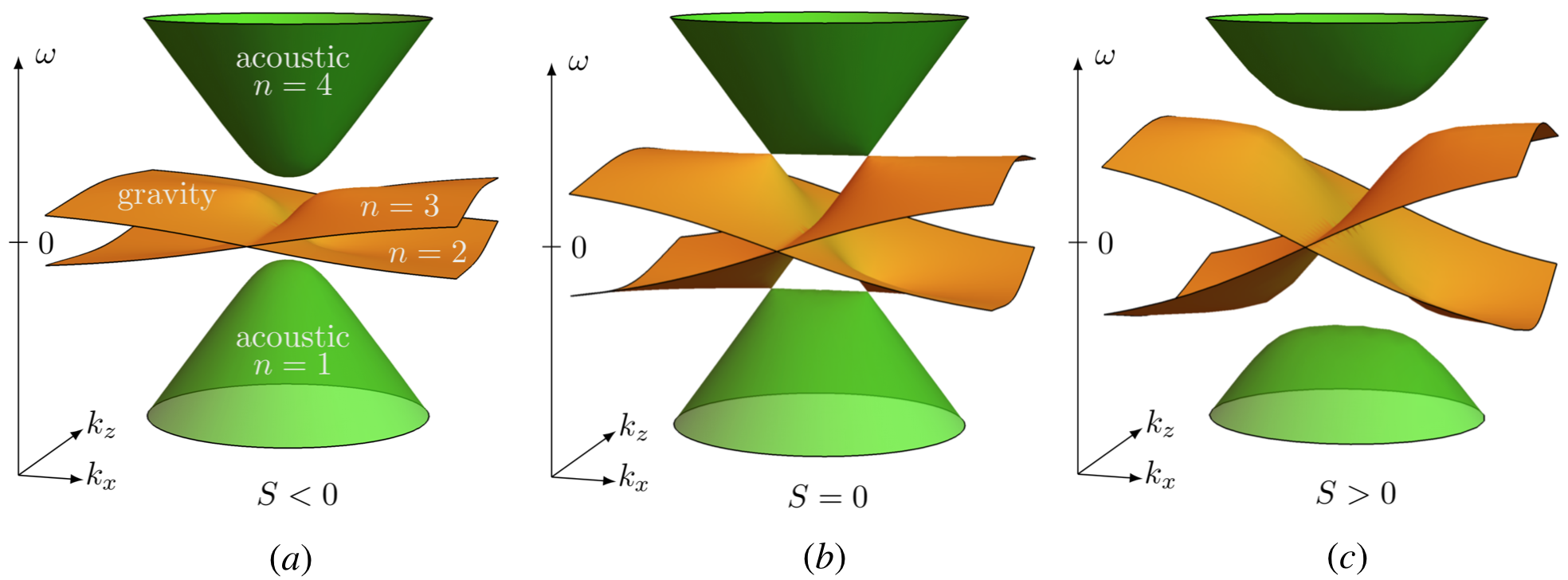}
\caption{\textcolor{black}{\textbf{Bulk dispersion relation for stratified compressible fluid, for different values of the stratification parameter $S$}}: (a) $S<0$;  (b) $S=0$; (c) $S>0$. \textcolor{black}{The index $n$ refers to the band number introduced in the text.}  The central internal-gravity wave bands (in orange, $n=2,3$) cross the acoustic wave bands (green, $n=1,4$) at four   \textcolor{black}{degeneracy} points  \textcolor{black}{if and only if $S=0$. In the vicinity of each degeneracy point, the dispersion relation has the shape of a tilted cone.}}
\label{fig:bulk} 
\end{figure}

A footprint of the broken mirror symmetry in $z$ direction is given by the presence of a frequency gap in the dispersion relation whenever $S\ne 0$ in unbounded geometry (see figure \ref{fig:bulk}).  The dispersion relation exhibits four bands $\omega^{(n)}$, corresponding to acoustic waves at high frequencies and internal gravity waves at low frequencies (see Methods).  
The gap between gravity and acoustic wave bands closes only when $S=0$ at $k_z=0$, that is to say when the mirror symmetry is recovered. In that case, the gap closes at four two-fold degeneracy points, where the dispersion has a conical shape reminiscent of tilted Dirac-like cones (see Methods). \textcolor{black}{The existence of degeneracy points is commonly associated with peculiar properties in the spectrum. Up to now, such properties have been overlooked in compressible stratified fluids, as only the case $S<0$ is usually discussed.}

\begin{figure}[h!]
\centering
\includegraphics[width=\textwidth]{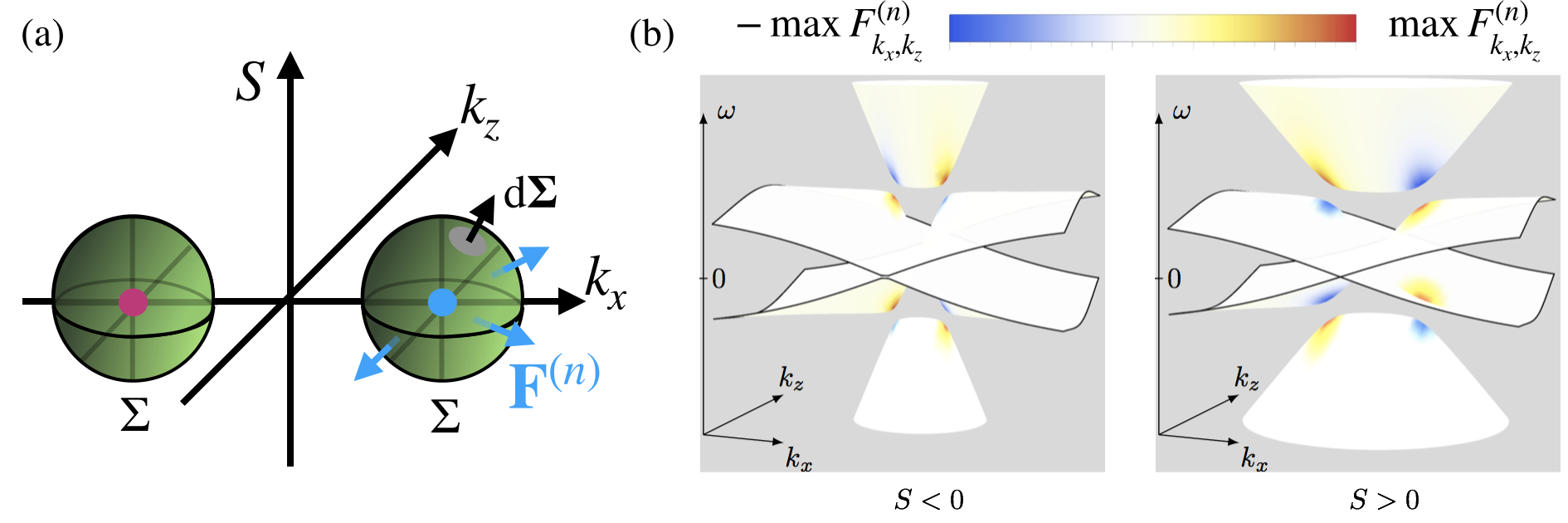}
\caption{\textbf{Berry curvature} (a) \textcolor{black}{Parameter space, with two degeneracy points (black and magenta). These degeneracy points induce a berry curvature $ \mathbf{F}^{(n)}$ in bundles of eigenmodes for each band $n$, similarly to  magnetic monopoles that induce  magnetic fields. The Chern number is the flux of this Berry curvature across the closed surface $\Sigma$.   (b)} The color code on the the dispersion relation is the value of the component $F_{k_x,k_z}^{(n)}$ of the Berry curvature for a prescribed value of $S$.} 
\label{fig:curvature} 
\end{figure}

Beyond the remarkable conical shape of the dispersion relation, degeneracy points are known to induce peculiar geometrical properties for the system eigenmodes %$\Psi^{(n)}$
in parameter space $(k_x,k_z,S)$\textcolor{black}{, see figure 2a}. The main point of this letter is to unveil these properties. They \textcolor{black}{can for instance be} revealed by the Berry curvature denoted $\mathbf{F}^{(n)}=(F_{k_x,k_z}^{(n)},\ F_{k_z,S}^{(n)},\ F_{S,k_x}^{(n)})$ (see Method). The $F_{k_x,k_z}^{(n)}$ component of the Berry curvature is shown for each band $n$ on the dispersion relation in figure \ref{fig:curvature}. Its amplitude is concentrated where the gaps are the smallest, as expected \cite{berry1984quantal}. More importantly, the amplitude changes sign with $S$. This reflects a topological property of the two-fold degeneracy points in $(k_x,k_z,S)$ space \cite{volovik2003universe}. 
\textcolor{black}{The Berry curvature induced by the two-fold degeneracy can be seen as an analog of a magnetic field induced by a monopole in parameter space. When integrated over a closed surface $\Sigma$ that encloses this degeneracy (as sketched in figure \ref{fig:curvature} (1a)), the flux of the Berry curvature is thus analogous to a charge that is quantized }
 $\int_\Sigma \mathbf{F}^{(n)} \cdot {\rm{d}} \boldsymbol{\Sigma} = 2\pi c^{(n)}$ where $c^{(n)}\in \mathbb{Z}$ is the first Chern number. For this reason, the degeneracies are said to carry topological charges $c^{(n)}$ \textcolor{black}{(sometimes called Berry monopoles by analogy with magnetic monopoles)} that are the source of the Berry curvature.

We find two degeneracy points in parameter space, at $(k_x,k_z,S)=(\pm N,0,0)$, leading to two sets of Chern numbers $c^{(n)}_{\pm}$. At the degeneracy point $k_x=N$, we obtain $c^{(n)}=-1$ for  acoustic wave bands ($n\in \{1,\ 4\}$), and $c^{(n)}=1$ for the internal-gravity wave bands ($n\in \{2,\ 3\}$), see Methods. These results bear strong similarities with the topological valley Hall effect in condensed matter \cite{xiao2007Valley, li2011topological, zhang2013valley}, with however subtle differences (see Methods). %The first one is that the Chern number of each band (in wave number space $(k_x,k_z)$ for fixed $S$) is ill-defined in the absence of a Brillouin zone, as it typically appears in a continuum model.  As a consequence, the base manifold is the open set $\mathbb{R}^2$ rather than the compact Brillouin torus, even though the integral  of the Berry curvature over $\mathbb{R}^2$ indeed vanishes for each band because of time-reversal symmetry. 

 The existence of these topological charges is crucial as it is related to the existence of a \textit{spectral flow} in the dispersion relation of the operator $\mathcal{H}_{\text{op}}$ that describes acoustic-gravity waves with varying stratification $S(z)$. This operator  is obtained from \eqref{eq:transf} after performing a Fourier transform on the basis $\ee^{i (\omega t -k_x x)}$, and can be formally deduced from the matrix $\mathcal{H}(k_x,k_z,S)$, by identifying $(k_x,k_z,S)\rightarrow (k_x,i\partial_z,S(z))$. The spectral flow means that some eigenvalues of $\mathcal{H}_{\text{op}}(k_x,\ii\partial_z,S(z))$ transit from the internal-gravity wave band to the acoustic wave band (or the other way around) when varying $k_x $, thus crossing a frequency gap in the vicinity of each degeneracy point.   
For this spectral flow to be non-zero -- or equivalently for the Chern number to be non-zero -- it is essential that $S$ changes sign. When $S$ is an increasing function of $z$, {and  given our choice of orientation for parameter space $(k_x,k_z,S)$, the algebraic number of modes $\mathcal{N}$ that flow to the band $n$ is given by $\mathcal{N}=c^{(n)}$ \cite{nakahara2003geometry,volovik2003universe,faure2000topological}. The sign of $\mathcal{N}$ is reversed when $S$ is a  decreasing function of $z$, and this can be generalized to non-monotonic profiles.} Moreover, the modes that transit are trapped around the altitude where $S$ vanishes. This is a common feature of topological domain-wall states encountered in condensed matter, from 1D chains \cite{jackiw1981solitons} to massless Dirac and Weyl fermions subjected to a magnetic field in 2D and 3D \cite{teo2010topological,liu2013chiral}.
The relation between the spectral flow of the operator $\mathcal{H}_{\text{op}}$ and the Chern numbers inferred from $\mathcal{H}(k_x,k_z,S)$ is rooted in the Atiyah-Singer index theorem \cite{nakahara2003geometry}. 
This deep connection has been successfully used to explain for instance the shape of molecular spectra \cite{faure2002topologically}, the chiral interface modes for generalized Dirac-like Hamiltonians in condensed matter \cite{fukui2012bulk,bal2018continuous}, and the existence of two eastward propagating equatorial waves \cite{delplace2017topological,faureexpose}. Based on this correspondence, we predict here a transition in the shape of the acoustic-gravity spectrum, controlled by the stratification profile $S(z)$.

\begin{figure}[h!]
   \centering
 \includegraphics[width=\textwidth]{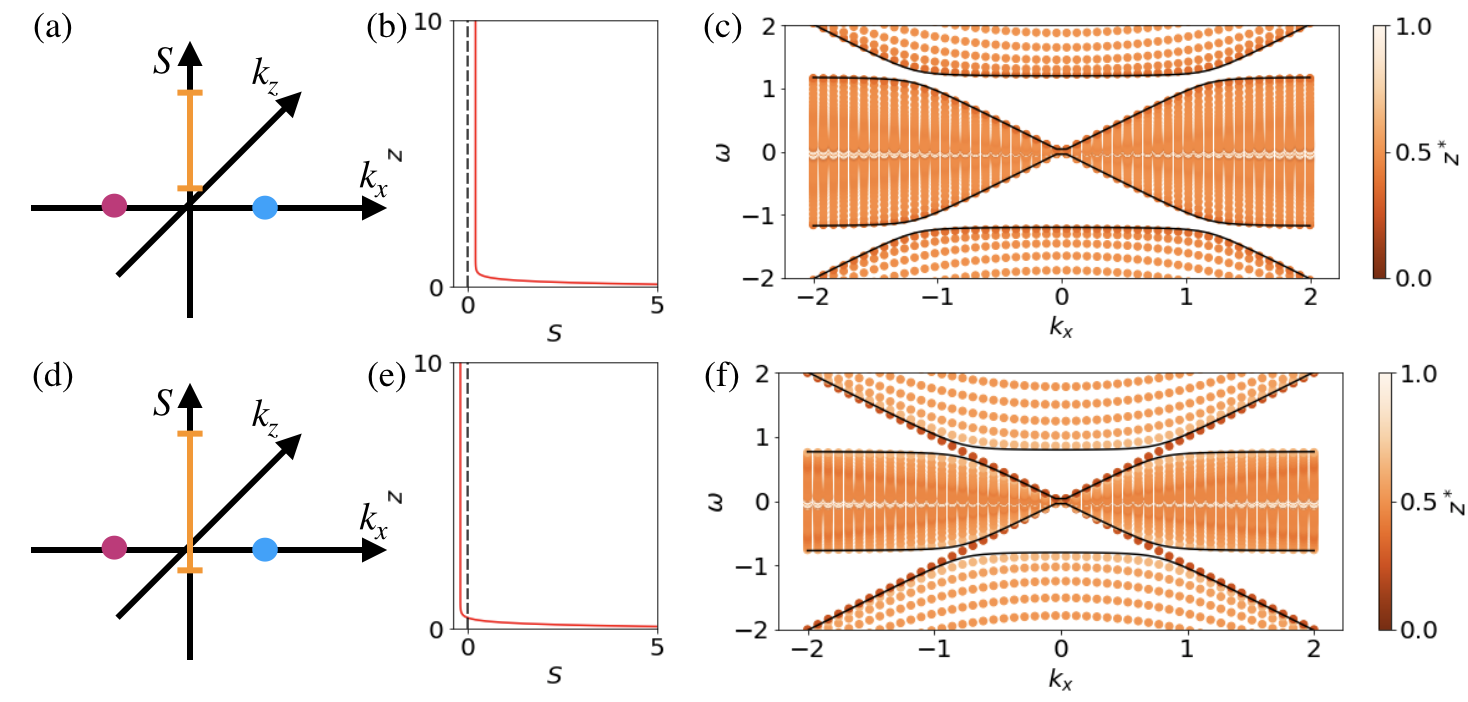}
    \caption{\textbf{Topological transition} in acoustic-gravity wave spectra, depending on the sign of $S_\infty$. \textcolor{black}{(a)  Parameter space for the \textit{bulk} problems dual to an \textit{interface} problem associated with a stratification profile shown in (b), where $S$ is always strictly positive.}  This profile is related to the buoyancy frequency profile $N(z)$ through $S=\sfrac{1}{2}(N^2-1)$. (c) Dispersion relation of the interface problem, showing acoustic and internal-gravity wave bands, and no spectral flow. Orange points correspond to the numerical spectrum, the intensity being proportional to localization of the waves at $z=z^* L_z$ (see method). The black line represents the analytical dispersion relation of the mode $m=0$ for the bulk problem $S=S_{\infty}$. \textcolor{black}{(d) Parameter space for the \textit{bulk} problems associated with the  stratification profile shown in (e), where $S$ changes sign.} (f) The corresponding dispersion relation exhibits a spectral flow consistent with the topological charges of the four degeneracy points (see text).}\label{fig:real}
\end{figure}

\textcolor{black}{As an application,} two stratification profiles of the form $S(z)=S_0e^{-z/z_0}+S_{\infty}$ and depicted in  figure \ref{fig:real} are considered. They only differ by the sign of $S_{\infty}$ \textcolor{black}{so that only the profile with $S_{\infty}<0$ changes sign with $z$. 
When $S_{\infty}>0$, then the stratification is always positive. It follows that one cannot define a surface $\Sigma$ in parameter space $(k_x,k_z,S)$ that wraps around the degeneracy point. The Chern number is thus zero. Accordingly, the spectrum of $\mathcal{H}_{\text{op}}$ is gapped.}
In contrast, when $S_{\infty}<0$, a spectral flow emerges and fills the frequency gap in the vicinity of the degeneracy points. This spectral flow is consistent with the Chern numbers computed above \textcolor{black}{for surfaces $\Sigma$ that enclose a degeneracy point}. Concretely, for positive wave numbers $k_x$, the acoustic wave bands gain one mode, while the internal-gravity wave bands loose one mode. This illustrates the existence of a topological transition when $S_{\infty}$ is varied.

%j ai commente cette phrase que je trouvais un peu alambiquee:
%\textcolor{black}{Although, the existence of these topological waves do not depend on the particular profile of $S(z)$ (provided it changes sign), the example we chose allows a comparison with the celebrated atmospheric Lamb waves. 
This particular example allows for a comparison with the celebrated atmospheric Lamb waves. 
Indeed, when the $e$-folding depth $z_0$ tends to zero, with $S_0$ sufficiently large, the profile of $S(z)$ tends to a constant $S_{\infty}$ in the bulk. For finite wave numbers, the spectra are found to be identical to the classical ones computed for  $S$ constant together with an impermeability condition $w=0$ at $z=0$ \cite{vallis2017atmospheric}. 
The similarity between the bottom intensified profile of $S(z)$ and the impermeability constraint is due to the fact that large stratification prevents vertical motion, thus mimicking the condition $w=0$ at $z=0$.
The modes that transit in this case correspond to the celebrated atmospheric Lamb waves \cite{lamb1911atmospheric}. %, that should not be confused with other Lamb waves encountered in elastic plates \cite{lamb1917waves}, and in particular in mechanical metamaterials \cite{wang2018topological}.
Atmospheric Lamb waves only exist when $S<0$ and may disappear whenever another boundary condition than the impermeability one is imposed \cite{iga2001transition}. In contrast, the topological \textcolor{black}{Lamb-like waves ccorresponding to the} spectral flow described here only depends on the zeros in the profile of $S$.

Our study shows that Lamb-like waves filling a frequency gap between acoustic and gravity waves can emerge even in the absence of a solid boundary, provided that the profile $S(z)$ changes sign \textcolor{black}{from a positive value to a negative value with increasing altitude. In that case the Lamb-like waves is trapped along the altitude where $S$ vanishes. 
We also predict with similar arguments an \textit{anti-Lamb} wave when $S$ changes sign from a negative value to a positive value with increasing altitude: this anti-Lamb mode would  transit from the  acoustic wave band at low wave number to the internal gravity wave band at large wavenumbers; to our knowledge, such waves have up to now never been observed.} 

 \textcolor{black}{Lamb-like waves could be observed in actual astrophysical and geophysical flows. The first issue is to find configurations where the stratification profile varies and crosses the critical value $N=g/c_s$ ($S=0$). The second issue is to avoid finite-size effects, due to the presence of actual boundaries (as a free surfaces or a solid ground). 
 These finite size effects would mix the Lamb-like wave existing without boundaries with other trapped waves associated with the actual flow boundaries. Indeed, the interactions between different  modes that transit from one waveband  to another can open gaps in the dispersion relation \cite{iga2001transition}. The trapping length scale of the Lamb-like wave close to the point where  $N=g/c_s$ is given by $c_s^2/g$. Finite size effects may be neglected at lowest order if this length is much larger than the flow depth $H$, and if the stratification threshold $N=g/c_s$ occurs in the domain bulk.}
 
\textcolor{black}{Stably stratified regions are found in planetary oceans atmospheres, at the base of the liquid core in planetary interiors, and in the radiative regions of stars (see supplementary materials). %at the base of the liquid core in planetary interiors \cite{souriau1991velocity}, and in the radiative regions of stars \cite{aerts2010asteroseismology}.
%Typical values of $c_s^2/gH$ and $N c_s/g$ for these different examples are recalled in supplementary material. 
Among all those cases, the most promising one for observing Lamb-like waves trapped away from a boundary is the radiative layers of stars, that are sufficiently deep, and sufficiently stratified to fulfill the two criterion given above. 
%To obtain quantitative predictions, it will be necessary to discuss the respective role of rotation, dissipation, mean flow, sphericity and spatial inhomegeneities on these topological waves.
 We stress that even the absence of Lamb-like wave in observations would provide useful information on the actual density profile  of the radiative layer, as this information can now be related to a constraint on the buoyancy frequency.}

%The main difficulty will be to find configuration with a suffiently small trapping length scale.
%%CE PARAGRAPHE PEUT ETRE ECOURTE AMON AVIS, AINSI QUE LA LISTE DES ARTICLES CITES
The present model is also of particular interest beyond geophysical applications. It shows a novel manifestation of topology in the context of acoustic waves that were first proposed in various artificial crystals \cite{acoustic_review}. %\cite{yang2015topological,mousavi2015topologically,khanikaev2015topologically,xia2015natureComm,he2016acoustic,fleury2016floquet}.% 
All previous examples of topological waves in classical systems with time-reversal symmetry have been designed in lattices enforcing specific symmetries to engineer  either Kramers degeneracies \cite{susstrunk2015observation,yves2017crystalline} (with $\mathbb{Z}_2$ invariants), or Dirac points \cite{pal2017edge,noh2018observation,qian2018theory,wang2018topological}. In the latter case, the topological phase thus achieved mimics the quantum valley Hall effect first proposed in bi-layer graphene \cite{xiao2007Valley,li2011topological,zhang2013valley}.  The topological spectral flow described here falls in this second class as it also relies on inversion (mirror) symmetry breaking. However, in stratified compressible flows, the presence of degeneracy points do not rely on  fine tuned symmetries, but naturally emerges at the resonance between acoustic waves and the intrinsic buoyancy frequency. This provides a useful platform to address the role of dissipation and nonlinearities on topological waves at a macroscopic level, obviating the need for specific metamaterials. 

\section*{Methods}

We consider in this paper the case of a stratified fluid in a vertical plane to simplify the presentation, but generalization of these results in the three dimensional case is straightforward.\\

\textbf{Stratification symmetry.} The dynamics \eqref{eq:transf} is left invariant by the transformation $(z,w,\theta,S) \rightarrow -(z,w,\theta,S)$ that we call stratification symmetry.  When $S=0$, mirror symmetry in the $z$-direction is restored. This occurs when $g=0$, as one can expect, but also when $N^2=g^2/c_s^2$ namely when the two intrinsic frequencies of the fluid match. By contrast, the system is always left invariant by two other discrete symmetries: mirror symmetry in the $x$ direction  $(x,u)\rightarrow-(x,u)$, and time-reversal symmetry $(t,u,w)\rightarrow-(t,u,w)$.\\

\textbf{Expressions of the symbol and the operator.} The \textit{symbol} $\mathcal{H}$ is obtained after performing a Fourier transform of \eqref{eq:transf} on the basis $\ee^{i (\omega t -k_x x - k_zz)}$, and considering $S$ (or equivalently $N$, assuming $g$ prescribed) as an external parameter. The dual \textit{operator}  $\mathcal{H}_{op}$  is obtained after performing a Fourier transform of \eqref{eq:transf} on the basis $\ee^{i (\omega t -k_x x)}$, considering now $S(z)$ as a given function of $z$. Their expression is 
\begin{align}
\mathcal{H}(k_x,k_z,S) = 
\begin{pmatrix}
0 & 0 & 0 & k_x \\
0 & 0 & \ii N & k_z - \ii S  \\
0 & -\ii N & 0      & 0\\
k_x  & k_z + \ii S  & 0   & 0
\end{pmatrix}, 
\quad
\mathcal{H}_{op}(k_x) = 
\begin{pmatrix}
0 & 0 & 0 & k_x \\
0 & 0 & \ii N(z) & \ii \partial_z - \ii S(z) \\
0 &- \ii N(z) & 0      & 0\\
k_x  & \ii \partial_z + \ii S(z)  & 0   & 0
\end{pmatrix} \label{eq:defH}
\end{align}
The stability condition of the fluid $N^2>0$ implies the condition $S>-1/2$ that needs to be fulfilled by the stratification parameter. If this condition is not satisfied, the symbol is not Hermitian.
\\

\textbf{Dispersion relation}  of acoustic gravity waves in unbounded geometries.  For each triplet of parameters $(k_x,k_z,S)$, the eigenfrequency and eigenmodes are obtained by solving $\omega \Psi = \mathcal{H} \Psi$ where $\Psi=(u,w,\theta,p)$, and where $\mathcal{H}$ is the Hermitian operator depending linearly on the parameters $(k_x,k_z,S)$ defined Eq. (\ref{eq:defH}). The dispersion relation $\omega_{\pm}^2=({K^2}/{2})\left[1\pm\left(1-(2Nk_x/K^2)^2\right)^{1/2}\right]$, with $K^2=k_x^2+k_z^2+N^2+S^2$, is plotted in figure \ref{fig:bulk} for three different values of $S$. This dispersion relation exhibits four bands $\omega^{(n)}$. High-frequency solutions $\omega^{(4)}=\omega_+$ and $\omega^{(1)}=-\omega_+$  are called acoustic waves, as their dispersion relation approaches $\omega_+^2=  (k_x^2+k_z^2)$ at sufficiently small scales (large wave numbers). Low-frequency solutions $\omega^{(3)}=\omega_-$ and $\omega^{(2)}=-\omega_-$ are called internal gravity waves, as their dispersion relation approach $\omega_-^2=N^2$ for sufficiently small scales. As expected from the analysis of symmetries, the dispersion relation is left invariant by  mirror symmetry in the $x$ direction ($k_x\rightarrow -k_x$) and time reversal symmetry ($\omega \rightarrow -\omega$). Interestingly, it is also left invariant by $k_z\rightarrow -k_z$, even though mirror  symmetry in $z$ is broken by the dynamics.
This property results from the combination of the two previous symmetries, together with a third one that is analogous to ``particle-hole'' symmetry in condensed matter physics. This third symmetry stems from the fact that the initial dynamical system describes real fields: each solution $\Psi(k_x,k_z,S)$ with eigenvalue $\omega$ has a partner $\Psi^*(-k_x,-k_z,S)$ with eigenvalue $-\omega$ (see supplementary materials).\\

\textbf{Degeneracy points.} When mirror symmetry is recovered, that is for $S=0$ and $k_z=0$, the dispersion relation can be interpreted in that case as the result of an horizontal sound wave $(u,p)$ of frequency $\omega = \pm  k_x$ coexisting with a stratified background media $(w,\theta)$ oscillating vertically at $\omega= \pm N$. This leads to four two-fold degeneracy points at $k_x^2= N^2=g^2/c_s^2$.
 The coupling between the fields $(u,p)$  and $(w,\theta)$ through the stratification parameter $S$ and vertical derivative leads to hybrid states separated by a band gap. This can be seen as a continuum counterpart of acoustic polaritons engineered recently in arrays of Helmholtz resonators \cite{kaina2015negative}.\\ 
 
 \textbf{Tilted Dirac-like cones.} The linear dispersion relation around the degeneracy points in ($k_x,k_z$)-space \textcolor{black}{has a conical shape (see Figure 1). This is referred to as a Dirac-like cone, similar to those emerging in graphene and other two-dimensional artificial systems \cite{montambaux2018} where the dispersion relation at low energy  is described by the massless Dirac equation. In the case of compressible-stratified fluids, these Dirac-like cones are tilted.  There has been a growing interest into tilted Dirac-cones in condensed matter, as they induce peculiar transport properties, and a classification has been proposed, depending on the intersection of the dispersion relation with the Fermi surface (here the plane with constant frequency including the degeneracy point) \cite{goerbig2008, trescher2015, zhang2017}. With its remarkable flat band along $k_x$, compressible stratified fluids realize unusual Dirac cones (dubbed as type III), that have so far only been observed in a honeycomb photonic lattice \cite{milicevic2018tilted}.}\\

\textbf{Berry Curvature} The Berry curvature in a 3D parameter space $(k_x,k_z,S)$ can be written as  $\mathbf{F}^{(n)}=(F_{k_x,k_z}^{(n)},\ F_{k_z,S}^{(n)},\ F_{S,k_x}^{(n)})$ whose components are
\begin{equation}
\label{eq:berry}
{F}_{p,p'}^{(n)}= i ( \partial_p  \Psi^{(n)*}_j\partial_{p'}\Psi^{(n)}_j - \partial_{p'}  \Psi^{(n)*}_j\partial_p\Psi^{(n)}_j )
\end{equation} where $\Psi_{j}^{(n)}$ is the $j^{\text{th}}$ component of the (normalized) $n^{\text{th}}$ eigenvector, $i^2=-1$, and $p$ and $p'$ are directions in parameter space $(k_x,k_z,S)$. The curvature \eqref{eq:berry} is gauge-invariant: it is independent of the choice of the eigenmode's phase. 

The Berry curvature is related to the existence of a non-zero Berry phase accumulated during the evolution of a wavepacket. In the context of acoustic-gravity waves, the emergence of non-zero Berry phases has been noted in Ref. \cite{budden1976phase,godin2015wentzel}, and has been related to the Hamiltonian structure of the underlying flow model \cite{vanneste1999wave}. However, the relation between this curvature, a broken discrete symmetry, and the existence of topological invariants carried by degeneracy points in parameter space, is new  in this context \cite{delplace2017topological}.\\

\textbf{Computation of the Chern numbers.} The two sets of Chern numbers $c^{(n)}_{\pm}$ assigned to degeneracy points $(k_x,k_z,S) = (\pm N,0,0)$ are related through mirror symmetry in $x$-direction: $c^{(n)}_{+}=-c^{(n)}_{-}$. Time reversal symmetry imposes that the Chern numbers of  bands with opposite frequencies are the same. Since the sum of Chern numbers at a given degeneracy point must vanish, the computation of the topological charges is eventually reduced down to the computation of a single Chern number  (see supplementary material).

The existence of two degeneracy points with opposite Chern numbers is reminiscent of the topological valley Hall effect \cite{xiao2007Valley, li2011topological, zhang2013valley}. However, we stress that in the case of acoustic-gravity waves, there is no underlying lattice structure.  The Chern number of each band (in wave number space $(k_x,k_z)$ for \textit{fixed} $S$) is ill-defined in the absence of a Brillouin zone: in a continuous medium, the base manifold is the open set $(k_x,k_z)\in \mathbb{R}^2$ rather than the compact Brillouin torus. The integral of the Berry curvature over the plane $(k_x,k_z)$ is not a Chern number, but it vanishes for each band because of time-reversal symmetry.\\ 

\textbf{Numerical computation of the spectra with arbitrary $S(z)$ profiles.} The spectra shown in figure 3 are computed with dedalus code \cite{dedalus}. In practice, for a given profile $S(z)$ in a domain $[0, L_z]$, the operator is projected on a Chebytchev basis, and the linear system is solved using the tau method. Boundary conditions are required at $z=0$ and $z=L_z$. We used $w=0$ at $z=0$, but we checked that other choice of the boundary conditions at this point did not affect the spectrum, given the choice of an exponential stratification profile. Indeed, this profile is equivalent to an effective condition $w=0$ around $z=0$. We follow the Iga prescription to avoid spurious eigenmodes at $z=L_z$ \cite{iga2001transition}; this amounts to impose the condition $p=0$ when $S>0$ and $w=0$ when $S<0$. This procedure allows us to identify the boundary $z=L_z$ with $z=\infty$, which can be checked by comparing the numerical spectra with the analytical spectra obtained in the case  of a flow taking place in the upper half space $z>0$, with $S=S_{\infty}$. For a given normalized eigenmode $\Psi(z)$, the localisation of the mode is computed as $z^*=\sfrac{1}{L_z}\int_0^{L_z} \Psi^*\cdot \Psi z\mathrm{d} z$.\\ 

\textbf{Data availability statement.} The data that support the plots within this paper and other findings of this study are available from the corresponding author upon request.

\bibliographystyle{naturemag}

\textbf{Acknowledgements} 

\textcolor{black}{We thank Louis-Alexandre Couston for his help with Dedalus code; Frederic Faure for providing useful insights on the index theorem; Thierry Alboussi\`ere, Isabelle Baraffe, Gilles Chabrier, Guillaume Laibe, Michael Le Bars for their input concerning potential geophysical and astrophysical applications; Leo Maas, Brad Marston and Nicolas Perez for useful comments on the manuscript. P. D. and A.V. were partly funded by  ANR-18-CE30-0002-01 during this work.}

\textbf{Author Contributions}
This paper emanates from the master project of  Manolis Perrot supervized by Pierre Delplace and Antoine Venaille. All the authors participated equally to the study. Pierre Delplace and Antoine Venaille wrote the paper. 

\textbf{Author Information} The authors declare no competing financial interests.

%%%%%%%%%%%%%Supplementary 

\pagebreak

\begin{center}
\textbf{\LARGE{Supplementary Material for:}}\\

\textbf{\LARGE  ``Topological transition in stratified fluids'', \textit{by Manolis Perrot, Pierre Delplace and Antoine Venaille}}
\end{center}

%%%%%%%%%% Merge with supplemental materials %%%%%%%%%%
%%%%%%%%%% Prefix a "S" to all equations, figures, tables and reset the counter %%%%%%%%%%

\setcounter{equation}{0}
\setcounter{figure}{0}
\setcounter{table}{0}
\setcounter{page}{1}
\makeatletter
\renewcommand{\theequation}{S\arabic{equation}}
\renewcommand{\thefigure}{S\arabic{figure}}

\section{Hermitian operator for acoustic-gravity waves}

The dynamics of an isothermal compressible stratified fluid around a state of rest can be written on the form of a Hermitian linear operator \cite{durran2010numerical}.  The general case, without hypothesis on the nature of the fluid, was discussed by \cite{iga2001transition}, in order to address the possible existence of Lamb waves in oceans. However, \cite{iga2001transition} did not write down  the linear operator on the Hermitian form (\ref{transf}). Here we show for completeness how to write down the linear dynamics in the form of an Hermitian operator. \textcolor{black}{The fact that such a transformation is possible can be traced back to the Hamiltonian structure of geophysical flow models, see e.g. \cite{salmon1998lectures, vanneste1999wave}.} 

We consider a compressible stratified fluid with a prescribed density profile $\rho_0(z)$ at rest. We restrict ourselves to the case of a non-rotating flow taking place in the plane $(x,z)$, with $z$ the vertical axis, in the direction opposite to the (constant) gravity field. 
Momentum and mass conservation equations write
\begin{eqnarray}
\rho D_t\mathbf{u}&=&-\mathbf{\nabla}p-\rho g\mathbf{e_z}\\
\partial_t\rho +\mathbf{\nabla}\cdot(\rho\mathbf{u})&=&0
\end{eqnarray}
where $\mathbf{u}=(u,w)$ is the velocity field, $\rho$ the fluid density, $p$ the pressure and  $D_t=\partial_t+(\mathbf{u}\cdot \nabla)$ the Lagrangian derivative.
Denoting $s$ the entropy field, pressure variation $p(s,\rho)$ is given by

\begin{align}
\dd p=\left(\frac{\partial p}{\partial s}\right)_{\rho} \dd s+\left(\frac{\partial p}{\partial \rho}\right)_{s} \dd \rho
\end{align}
Assuming adiabatic particle displacements ($\dd s=0$) leads to 
\begin{equation}
D_t\rho=\frac{1}{c_s^2}D_tp
\end{equation}
where $c_s^2=\left(\frac{\partial p}{\partial \rho}\right)_{s}$ is the square of the sound speed.\\
In basic state $(\rho_0(z)$, $p_0(z))$, the fluid is at rest and hydrostatically balanced
\begin{align}
    \frac{\dd p_0}{\dd z}=-\rho_0g. \label{hydro}
\end{align}
We will consider small disturbances $\mathbf{u}'$, $\rho'$, $p'$ of this basic state
\begin{align}
\mathbf{u}=\mathbf{0}+\mathbf{u'}(x,z,t) ,\ \rho=\rho_0(z)+\rho '(x,z,t),\ p=p_0(z)+p'(x,z,t)\\
u',w'\ll c_s,\ \rho'\ll\rho_0,\ p'\ll p_0. 
\end{align}

Using \ref{hydro} to express advection of pressure leads to

\begin{subequations}
\begin{empheq}[left=\empheqlbrace]{align}
\rho_0\partial_t\mathbf{u'}&=-\nabla p'-\rho'g\mathbf{e_z}\\
\partial_t\rho'+w'\partial_z\rho_0+\rho_0\nabla \cdot \mathbf{u'}&=0\\
\partial_t\rho'+w'\partial_z\rho_0&=\frac{1}{c_s^2}(\partial_tp'-w'\rho_0g)
\end{empheq}
\end{subequations}
We use the following change of variables
\begin{align}
\label{chgtrho}
\tilde{u}=u'\rho_0^{1/2},\ \tilde{w}=w'\rho_0^{1/2},\ \tilde{p}=p'\rho_0^{-1/2},\ \tilde{\rho}=\rho'\rho_0^{-1/2}
\end{align}
that leads to 
\begin{subequations}
\begin{empheq}[left=\empheqlbrace]{align}
\partial_t\tilde{u}&=-\partial_x\tilde{p}\\
\partial_t\tilde{w}&=-\partial_z\tilde{p}-\frac{1}{2}\frac{\partial_z\rho_0}{\rho_0}\tilde{p}-\tilde{\rho}g\\
\partial_t\tilde{p}&=g\tilde{w}-c_s^2\nabla \cdot\mathbf{\tilde{u}}+\frac{1}{2}c_s^2\frac{\partial_z\rho_0}{\rho_0}\tilde{w}\\
\partial_t\tilde{\rho}&=-\tilde{w}\frac{\partial_z\rho_0}{\rho_0}+\frac{1}{c_s^2}(\partial_t\tilde{p}-\tilde{w}g)
\end{empheq}
\end{subequations}
The density field is substituted by
$\tilde{\theta}=\tilde{\rho}-\frac{1}{c_s^2}\tilde{p}$. This leads to the following set of equations:
\begin{subequations}
\begin{empheq}[left=\empheqlbrace]{align}
\partial_t\tilde{u}&=-\partial_x\tilde{p}\\
\partial_t\tilde{w}&=-\partial_z\tilde{p}-\left(\frac{1}{2}\frac{\partial_z\rho_0}{\rho_0}+\frac{g}{c_s^2}\right)\tilde{p}-\tilde{\theta}g\\
\partial_t\tilde{p}&=\left(g+\frac{1}{2}c_s^2\frac{\partial_z\rho_0}{\rho_0}\right)\tilde{w}-c_s^2\nabla \cdot\mathbf{\tilde{u}}\\
\partial_t\tilde{\theta}&=-\left(\frac{\partial_z\rho_0}{\rho_0}+\frac{g}{c_s^2}\right)\tilde{w}
\end{empheq}
\end{subequations}
Let us introduce the square of the buoyancy frequency
\begin{equation}
N^2=-g\frac{\partial_z\rho_0}{\rho_0}-\frac{g^2}{c_s^2}.
\end{equation}
 Stable configurations correspond to $N^2>0$. In that case $N$ is interpreted as an intrinsic frequency of the system, referred to as the Brunt-V\"ais\"al\"a frequency, or the buoyancy frequency. We also introduce another frequency of the system
\begin{equation}
S=\frac{1}{2}\left(\frac{N^2c_s}{g}-\frac{g}{c_s}\right).
\end{equation}
Assuming $c_s$ constant, and denoting $\bar{\theta}=\tilde{\theta}N/g$, 
\begin{align}\label{transf}
\partial_t\left(\begin{array}{c}
\tilde{u}\\ \tilde{w} \\ \bar{\theta} \\ \bar{p}
\end{array}\right)=
\left(\begin{array}{cccc}
0 & 0 &0&-c_s\partial_x\\
0&0&-N&S-c_s\partial_z\\
0&N&0&0\\
-c_s\partial_x&-S-c_s\partial_z&0&0
\end{array}\right)\left(\begin{array}{c}
\tilde{u}\\ \tilde{w} \\ \bar{\theta} \\ \bar{p}
\end{array}\right)
\end{align}
This is the operator studied in the main text, where the symbolds " $\tilde{}$ " and " $\bar{}$ " are dropped

\section{Stratification parameter in isothermal atmospheres}
In non-isothermal atmospheres, non-constant stratification profiles with $S<0$ and $S>0$ can occur. However in most textbooks, Lamb waves are derived from the case of an isothermal atmosphere. Here we show that in that case thermodynamical constraints impose $S<0$.  \\
Let's consider that the atmosphere is described as an ideal gas at a constant temperature $T_0$. The state equation $p_0(z)=\rho_0(z)R T_0$ holds, where $R$ is the specific gas constant. At rest, the atmosphere is hydrostatically balanced, so $\frac{\dd p_0}{\dd z}=-\rho_0(z)g$. Combining the two last equations yields to the density profile of the isothermal atmosphere : \begin{align}
    \rho_0(z)=\rho_0(z=0)\mathrm{e}^{-\frac{gz}{RT_0}}
\end{align} 
Now recall that the stratification parameter is given by $S=1/2(-c_s\partial_z\rho_0/\rho_0-2g/c_s)$ and that the sound speed in an ideal gas is $c_s=\sqrt{\gamma RT_0}$, where $\gamma$ is the heat capacity ratio. This leads to 
\begin{align}
    S=\frac{g}{2c_s}(\gamma-2)
\end{align}
which is always negative. Indeed in ideal gases, $\gamma = 1+2/f$, where $f\ge 3$ is the number of degrees of freedom for a single gas molecule.

\section{Symmetries of bulk waves}
The symmetries of the partial differential equations describing acoustic-gravity waves are given in the main text. These symmetries can now be translated in Fourier space $(t,x,z)\rightarrow (\omega,k_x,k_z)$, as symmetries of the operator $\mathcal{H}(k_x,k_z,S,N)$ defined in Eq. (4) of the method section. When doing so, several arbitrary choices are possible, each of them being compatible when all symmetries are taken into account. 

As an example, consider time reversal symmetry $t\rightarrow -t$ that imposes $u\rightarrow -u$. Let us then decompose the velocity in Fourier space as 
\begin{equation}
u(t,\mathbf{x})=\int \mathrm{d} \omega \mathrm{d} \mathbf{k} e^{i\omega t-i\mathbf{k}\cdot \mathbf{x}} \hat{u}(\omega,\mathbf{k}) .
\end{equation}
 Since $u(t,\mathbf{x})$ is a real-valued vector field, this decomposition imposes $\hat{u}(\omega,\mathbf{k}) = \hat{u}^*(-\omega,\mathbf{-k})$. At the level of the operator $\mathcal{H}(k_x,k_z,S,N)$, this \textit{real field condition} leads to $\mathcal{H}(k_x,k_z,S,N)=-\mathcal{H}(-k_x,-k_z,S,N)^*$, that is the analog of particle-hole symmetry in condensed matter. As a consequence, the frequency spectrum has the symmetry $\omega(k_x,k_z) \rightarrow -\omega(-k_x,-k_z)$.

Time-reversal symmetry  $u(t,\mathbf{x})\rightarrow -u(-t,\mathbf{x})$ can then be satisfied in two different ways:
\begin{itemize}
\item (i) $\omega\rightarrow -\omega$ and $\hat{u}(\omega,\mathbf{k}) \rightarrow\hat{u}(-\omega,\mathbf{k})  $,
\item (ii)  $i\rightarrow -i$, $\mathbf{k}\rightarrow -\mathbf{k}$ and $\hat{u}(\omega,\mathbf{k}) \rightarrow\hat{u}(\omega,-\mathbf{k})  $.
\end{itemize}

Here we refer to choice (i) as time reversal symmetry for the Fourier modes, because $\omega$ has the dimension of the inverse of a time. Choice (ii) coincides  with the standard time reversal symmetry in quantum mechanics, where $\mathbf{k}$ is the momentum. In that case there exists an anti-unitary operator that commutes with  $\mathcal{H}(k_x,k_z,S,N)$ for $\mathbf{k}=0$. In contrast, choice (i) leads to the existence of a unitary operator that anti-commutes with $\mathcal{H}(k_x,k_z,S,N)$. It follows that the frequency spectrum has the symmetry $\omega(k_x,k_z) \rightarrow -\omega(k_x,k_z)$. This usually refers to chiral or sublattice or bi-partite symmetry in condensed matter. Choice (ii) is automatically satisfied by combining the real field condition with choice (i).

All the symmetries of the physical system can then be formally rephrased in Fourier space by introducing a unitary operator for each of them. This draws an interpretation of $\mathcal{H}(k_x,k_z,S,N)$ as a classical counterpart of a quantum Hamiltonian: 

\begin{table}[h!]
\begin{tabular}{|l|l|l|}
\hline
  Symmetry                   &  Operator                                 &  Commutation Relation \\
  \hline
  \hline
  Mirror in $x$ : $(x,u) \rightarrow -(x,u)$   &  $\mathcal{M} =\text{diag}(-1,1,1,1)$  & $\mathcal{M} \mathcal{H}(k_x,k_z,S,N) \mathcal{M}^{-1} = \mathcal{H}(-k_x,k_z,S,N)$ \\
  \hline
Time reversal:       $(t,u,w) \rightarrow -(t,u,w)$   &  $\mathcal{T} = \text{diag}(1,1,-1,-1)$  & $\mathcal{T} \mathcal{H}(k_x,k_z,S,N) \mathcal{T}^{-1} = -\mathcal{H}(k_x,k_z,S,N)$ \\
  \hline
 Stratification:  $(z,w,\theta,S) \rightarrow -(z,w,\theta,S) $   &  $\mathcal{S} = \text{diag}(1,-1,-1,1)$  & $\mathcal{S} \mathcal{H}(k_x,k_z,S,N) \mathcal{S}^{-1} = \mathcal{H}(k_x,-k_z,-S,N)$ \\
  \hline
Real fields  $(u,w,\theta,p) \in \mathbb{R}^4 $   & complex conjugation $\mathcal{K}$ & $ \mathcal{K}\mathcal{H}(k_x,k_z,S,N)\mathcal{K}^{-1} =-\mathcal{H} (-k_x,-k_z,S,N)$ \\
  \hline
\end{tabular}
\caption{Symmetries of the linear operator projected in Fourier space.}
\label{table}
\end{table}

\section{Computation of the Chern number}

Here we show how to compute the Chern numbers associated with degeneracy points in the spectrum of the symbol $\HH(k_x,k_z,S)$ defined Eq. (4) in method section. This $4$ by $4$ matrix can be written as
\begin{align}\HH=\begin{pmatrix}
0&h\\
h^{\dagger}&0
\end{pmatrix},\ \text{where}\ h=\begin{pmatrix}
0&k_x\\
    \ii N&k_z-\ii S\\
\end{pmatrix} \ .
\end{align}

The study of the spectrum of $\HH$ is then conveniently reduced down to the study of the spectrum of $hh^{\dagger}$, provided that the eigenvalues of $\HH$ are non zero. Indeed each solution of $hh^{\dagger}\phi=\alpha\phi$ corresponds to two solutions of $\omega\Psi=\HH\Psi$, with $\omega=\sqrt{\alpha}$ and $\Psi=(\phi,(h^{\dagger}\phi/\omega))$. This reduction procedure is classically used in textbooks on compressible stratified fluids, when writing the dynamics as a second order in time linear PDE for $u$ and $w$ \cite{vallis2017atmospheric,iga2001transition}. One can check that the Berry curvature generated by $\phi$ is the same as the Berry curvature generated by $\Psi$.
The matrix $hh^{\dagger}$ is expressed as
\begin{align}
    hh^{\dagger}=\frac{k_x^2+k_z^2+N^2+S^2}{2}I_2+\g\cdot\sigma\ , \ \g=\begin{pmatrix}
    k_xk_z\\-k_xS\\ \frac{k_x^2-k_z^2-N^2-S^2}{2}
    \end{pmatrix}
\end{align}
where $\sigma=(\sigma_x,\sigma_y,\sigma_z)$ is the vector of Pauli matrices. In that form, eigenvalues of $hh^{\dagger}$ are given by 
\begin{align}
\alpha^{(n)}=\frac{k_x^2+k_z^2+N^2+S^2}{2}+(-1)^n\|\mathbf{g}\|, \ n=1,2 .
\end{align} 
In that 2 by 2 case, the wave bands $n=1$ and $n=2$ correspond to gravity waves and acoustic waves, respectively. \textcolor{black}{Note that the degeneracy point of the gravity bands at $k_x=0$ has disappeared}. 

From now on, we consider $N^2=(2 S+1)$.  
The degeneracy points $\mathbf{p}_{\pm}=(k_x=\pm 1,k_z=0,S=0)$ are easily identified as they correspond to nullification points of $\|\mathbf{g}\|$. 
A classical calculation leads to 
\begin{align}
   \frac{1}{2\pi}F^{(n)}_{p,p'}=(-1)^{(n)}\frac{1}{4\pi}\frac{\g}{\|\g\|^3}\cdot\left(\partial_p\g\times\partial_{p'} \g\right) ,
\end{align}
see e.g. \cite{fruchart2013introduction}. This allows us to identify the topological charge of the lower (internal-gravity) band $c^{(1)}_{\pm}$ carried by a degeneracy point $\mathbf{p}_{\pm}$ with the index  of the vector field $\g$ at $\mathbf{p}_{\pm}$. Finally this index equals $\mathrm{sign}(\mathrm{det}(\dd_{\mathbf{p}_{\pm}}\g))$, where $\dd_{\mathbf{p}_{\pm}}\g$ is the Jacobian matrix (the matrix of partial differentials). We find

\begin{align}
    \dd_{\mathbf{p}_{+}}\g=\begin{pmatrix}
0&1&0\\
0&0&-1\\
1&0 &-1
\end{pmatrix}
\end{align}
so the topological charge of the gravity wave band is $c^{(1)}_{+}=1$ at $k_x=1$.  Note that in the 4 by 4 problem, the gravity wave and acoustic wave bands are indexed by  $n\in\{2,3\} $ and   $n\in\{1,4\} $, respectively. The topological charges of the degeneracy points for these bands are deduced from the 2 by 2 case, using symmetries of the problem:
\begin{align}
  (c_+^{(1)},  c_+^{(2)},  c_+^{(3)},  c_+^{(4)}) = (-1,\ 1,\ 1, -1)\ &\text{at } k_x=1 , \\
  (c_-^{(1)},  c_-^{(2)},  c_-^{(3)},  c_-^{(4)}) = (1,\ -1,\ -1, 1)\ &\text{at } k_x=-1 .
\end{align}

\section{\textcolor{black}
{Parameters for astrophysical and geophysical flows}}

\color{black}
{The existence of topological Lamb-like waves is guaranteed in a domain of fluid where $S$ changes sign, that is $Nc_s/g \sim 1$. Moreover, since these waves are trapped over a characteristic length $c^2/g$, they would be thus more easily identifiable if they are not coupled to the standard boundary Lamb wave. Thus, one should ideally look for a system with a sufficiently  large flow depth $H$, i.e. $c^2/gH << 1$. Different possibilities are listed below.}

\begin{table}[h!]
\centering
\color{black}
\begin{tabular}{|l|c|c|c|c||c||c||}
    \cline{2-7}
\multicolumn{1}{l|}{}  & $g$ (m.s$^{-2}$) & $c$ (m.s$^{-1}$) & $N$ (s$^{-1}$) & $H$ (m) &$\frac{c^2}{gH}$ & $\frac{Nc}{g}$\\
    \hline
    \textbf{Earth's atmosphere}$^{(a)}$ &  $10$  & $3.\  10^2$  & $10^{-2}$ & $ 8. \  10^4$  & $8$ &  $<1$\\
    \hline 
    \textbf{Earth's ocean thermocline}$^{(b)}$ &  $10$  & $1.5 \ 10^3$  & $10^{-2}$ & $ 10^3$  & $200$ & $\sim 1$  \\
     \hline
   \textbf{Earth's liquid core} (base)$^{(c)}$&   $4$  & $10^4$  & unknown & $2.\ 10^5$  &  100 & unknown \\
    \hline
   \textbf{Solar radiative zone}$^{(d)}$ &   $10^3$  & $3.  10^5$  & $ 3.\  10^{-3}$ & $4.\  10^8$  &  0.25 & $\sim 1$ \\
    \hline
  \end{tabular}
 
\end{table}

\begin{enumerate}[(a)]
\item  The depth $H$ is the height of the stratopause (top of the stratosphere). The stratification is twice bigger in the stratosphere ($z>10$ km) than in the troposphere $z<10$ km). As a whole, Earth's atmosphere can at first order be considered as isothermal gas, for which we know that $Nc/g<1$, even when $Nc/g \sim 1$ \cite{vallis2017atmospheric}.

 \item The Earth's oceans have everywhere a vertical structure  with a thermocline ($1$ km depth) above the abyss ($4$ km depth). The  stratification of the thermocline is an order of magnitude larger in the upper part of the thermocline  than in the deep abyss where $N\sim 10^{-3}$ s$^{-1}$ \cite{vallis2017atmospheric}. 
 
 \item  The existence of a stably stratified at the base of the Earth's liquid core is inferred in ref. \cite{souriau1991velocity}. \\
  
\item  The radiative region of the sun is a stably stratified layer located at $r=0.2R_{\astrosun}$ and $r=0.7 R_{\astrosun}$   with $R_{\astrosun}$ the sun radius; the depth is thus estimated as $H=0.5 R_{\astrosun}$. Parameters given in the table are estimated around  $r\approx 0.4 R_{\astrosun}$, using ref \cite{aerts2010asteroseismology}. At this location, there is almost a band crossing between gravity waves (called $g$ modes) and acoustic waves (called $p$ modes). 

\end{enumerate}

We have seen that the criterion $N=g/c_s$ is never satisfied in the Earth atmosphere, assuming that it is not far from isothermal (or more weakly stratified). The same conclusion probably holds for other planetary atmospheres, including giant planets, and for stellar atmospheres. 

The criterion $N=g/c_s$ is marginally satisfied in the upper thermocline, and it is not hard to imagine other possible climates  where this criterion would be fulfilled in the interior. However, the typical depth $c_s^2/g$ is much larger than the ocean depth, and one can not avoid a discussion on the effect of interactions with surface waves on the global spectrum. One could look for other planets with deep oceans, as in Ganymed, where  $H=100$ km. However, this depth remains much larger than the ratio $c_s^2/g$ in this case.

Among the different examples listed in the table, the best candidate for potential observations of Lamb-like waves trapped at the critical stratification $N=g/c_s$ away from boundaries is the radiative zone of the sun, as it is sufficiently deep, and sufficiently stratified. Even if the criterion $c_s^2/gH\ll 1$ is marginally satisfied in the solar radiative zone, a discussion of finite size effects may be necessary. If fact, the observations of surface waves (called $f$-modes) that bridge the gap between acoustic modes and gravity modes shows that the actual boundaries have to be taken into account to describe the global shape of the spectra. More generally we expect that the criteria for Lamb-like waves will be satisfied in the radiative zone of most stars.  Massive stars and red giants would be of particular interest, as their radiative zone is located outside the convective layer, which makes easier observations of possible waves in these layers.

 \end{document}